\documentclass[sn-mathphys,Numbered]{sn-jnl}
\usepackage{graphicx}%
\usepackage{multirow}%
\usepackage{amsmath,amssymb,amsfonts}%
\usepackage{amsthm}%
\usepackage{mathrsfs}%
\usepackage[title]{appendix}%
\usepackage{xcolor}%
\usepackage{textcomp}%
\usepackage{manyfoot}%
\usepackage{booktabs}%
\usepackage{algorithm}%
\usepackage{algorithmicx}%
\usepackage{algpseudocode}%
\usepackage{listings}%
\usepackage[justification=centering]{caption}
\usepackage{tabularx, booktabs, threeparttable, caption}
\theoremstyle{thmstyleone}%

\theoremstyle{thmstyletwo}%
\theoremstyle{thmstylethree}%
\begin{document}

\title[Quantum]{A Quantum Convolutional Neural Network Approach for Object Detection and Classification}

\author*[1]{\fnm{Gowri Namratha} \sur{Meedinti}}\email{gowri.namratha2019@vitstudent.ac.in}

\author*[1]{\fnm{Kandukuri Sai} \sur{Srirekha}}\email{kandukurisai.srirekha2019@vitstudent.ac.in}

\author[1]{\fnm{Radhakrishnan} \sur{Delhibabu}}\email{rdelhibabu@vit.ac.in}
\equalcont{These authors contributed equally to this work.}

\affil[1]{\orgdiv{School of Computing Science and Engineering (SCOPE)}, \orgname{Vellore Institute of Technology}, \orgaddress{\city{Vellore}, \postcode{632014},\country{India}}}


\abstract{This paper presents a comprehensive evaluation of the potential of Quantum Convolutional Neural Networks (QCNNs) in comparison to classical Convolutional Neural Networks (CNNs) and Artificial / Classical Neural Network (ANN) models. With the increasing amount of data, utilizing computing methods like CNN in real-time has become challenging. QCNNs overcome this challenge by utilizing qubits to represent data in a quantum environment and applying CNN structures to quantum computers. The time and accuracy of QCNNs are compared with classical CNNs and ANN models under different conditions such as batch size and input size. The maximum complexity level that QCNNs can handle in terms of these parameters is also investigated. The analysis shows that QCNNs have the potential to outperform both classical CNNs and ANN models in terms of accuracy and efficiency for certain applications, demonstrating their promise as a powerful tool in the field of machine learning.}

\maketitle

\section{Introduction}\label{sec1}

In recent years, there has been a significant increase in investment in the field of quantum computing, with the aim of leveraging its principles to solve problems that are intractable using traditional computing techniques. The intersection of quantum computing and deep learning is of particular interest, as both fields have seen significant growth in recent years.  Researchers such as Garg and Ramakrishnan \cite{lib_1} have highlighted the potential of quantum computing to revolutionize current techniques in areas such as security and network communication. The application of quantum computing principles to deep learning models has the potential to significantly enhance their performance and enable the solution of classically intractable problems. As such, there has been a growing interest in the exploration of the possibilities at the intersection of these two fields, commonly referred to as Quantum deep learning.

The classification outcome is obtained by utilizing the fully connected layer after the data size has been effectively reduced by multiple applications of these layers. To achieve optimal results, the discrepancy between the acquired label and the actual label can be employed to train the model using optimization techniques such as gradient descent. In recent years, several studies have been conducted that combine the principles of quantum computing and the CNN model to solve real-world problems that are otherwise intractable using conventional machine learning techniques through the use of Quantum Convolutional Neural Networks (QCNN). There exists an approach for efficiently solving quantum physics problems by incorporating the CNN structure into a quantum system, as well as a methodology for enhancing performance by incorporating quantum principles into problems previously solved by CNN.

\section{Background}\label{sec2}
\subsection{Convolutional Neural Network}
Convolutional Neural Networks (CCNNs) are a subclass of artificial neural networks that are widely utilized in image recognition and audio processing tasks. They possess the ability to identify specific features and patterns in a given input, making them a powerful tool in the field of computer vision. The ability to identify features is achieved by using two types of layers in a CCNN: the convolutional layer and pooling layer (Figure \ref{fig2})

The convolutional layer applies a set of filters or kernels to an input image, resulting in a feature map that represents the input image with the filters applied. These layers can be stacked to create more complex models, which can learn more intricate features from images. The pooling layer, on the other hand, reduces the spatial size of the input, making it easier to process and requiring less memory. They also help to reduce the number of parameters and speed up the training process. Two main types of pooling are used: max pooling and average pooling. Max pooling takes the maximum value from each feature map, while average pooling takes the average value. Pooling layers are typically used after convolutional layers to reduce the size of the input before it is fed into a fully connected layer.

Fully connected layers are one of the most basic types of layers in a CNN, where each neuron is connected to every other neuron in the previous layer. They are typically used towards the end of a CNN, when the goal is to take the
 
features learned by the previous layers and use them to make predictions. For example, if a CNN is used to classify images of animals, the final fully connected layer might take the features learned by the previous layers and use them to classify an image as containing a dog, cat, bird, etc.

\begin{figure}[h]
\begin{center}
    \includegraphics[scale=0.3]{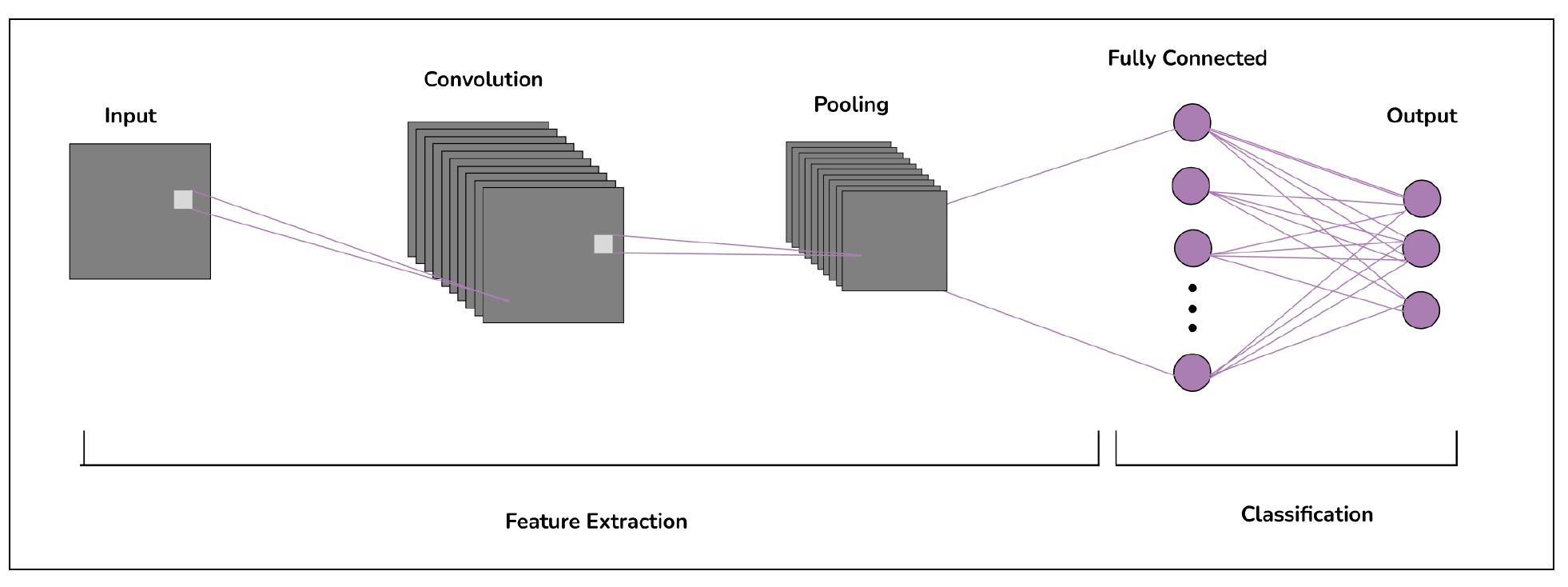}
        \caption{Convolutional Neural Networks \label{fig2}}
    \end{center}
\end{figure} 

Several studies have been conducted to improve the performance of CCNNs. For example, Saad Albawi et al.  \cite{lib_2} discussed the layers of the Convolutional Neural Network in depth and found that as the number of layers and parametersincrease, the time and complexity significantly increase for training and testing the model. The use of convolutional neural networks (CNNs) in image analysis tasks has been extensively studied in recent literature. Keiron O'Shea et al. in their study \cite{lib_3} discuss the advantages of CNNs over traditional artificial neural networks (ANNs) and the best ways to structure a network for image analysis tasks. The authors highlight that CNNs exploit the knowledge of the specific input used, but also note that they are resource- heavy algorithms, particularly when dealing with large images. Shabeer Basha et al. in \cite{lib_4} investigate the relationship between fully connected layers and CNN architectures, and the impact of deeper/shallower architectures on CNN performance. The authors conclude that shallow CNNs perform better with wider datasets, while deeper CNNs are more suitable for deeper datasets. This serves as an advantage for the deep learning community as it allows for the selection of the appropriate model for higher precision and accuracy on a given dataset.

In \cite{lib_5}, Sakshi Indolia et al. highlight the architectures and learning algorithms used for CNNs. They mention that the GoogleNet architecture, while reducing the budget and number of trainable parameters, also increases the risk of overfitting as the network size increases. Youhui Tian in \cite{lib_6} presents a new CNN algorithm that aims to increase convergence speed and recognition accuracy by incorporating a recurrent neural network and a residual module called ShortCut3-ResNet. This ultra-lightweight network structure reduces the number of parameters, making the algorithm more diverse in feature extraction and improving test accuracy.

Shyava Tripathi et al. in \cite{lib_7} focus on the real-time implementation of image recognition with a low complexity and good classification accuracy for a dataset of 200 classes. The authors suggest that there is scope for improvement in increasing the number of classes to 1000 and focusing on feature extraction rather than raw input images. In \cite{lib_8}, Rahul Chauhan et al. implement the algorithm on MNIST and CIFAR-10 datasets, achieving 99.6 percent and 80.17 percent accuracy, respectively. The authors suggest that the accuracy on the training set can be improved by adding more hidden layers.

Deepika Jaiswal et al. in \cite{lib_9} implement the algorithm against various standard data sets and measure the performance based on mean square error and classification accuracy. The classification accuracy for some datasets reaches 99 percent and 90 percent, but for others, such as large aerial images, the accuracy is in the 60s and 70s, indicating scope for improvement. In \cite{lib_10}, Neha Sharma et al. conduct an empirical analysis of popular neural networks like AlexNets, GoogleNet, and ResNet50 against 5 standard image data sets. The authors find that 27 layers are insufficient to classify the datasets and that the more layers, the higher the accuracy in prediction. In this case, the highest accuracy was achieved at 147-177 layers, which is not suitable for training on a normal desktop. However, once trained, the model can be used in a wide number of applications due to its flexibility.

Finally, in \cite{lib_11}, Shuying Liu and Weihong Deng aim to prove that deep CNNs can be used to fit small datasets with simple modifications without severe overfitting. The authors conclude that on large enough images, batch normalization on a very deep model will give comparable accuracy to shallow models. However, they also note that there is still scope for improvement in both methods, suggesting that deep models can be used for small datasets once overfitting is addressed and better accuracy is achieved.

Overall, the literature suggests that CNNs are powerful tools for image analysis tasks and that various architectures and modifications can be used to improve performance on different datasets. Convolutional Neural Networks (CNNs) have been widely used in image analysis tasks and have shown to be effective in achieving high accuracy and precision. However, there are still challenges and areas for improvement, such as memory allocation for large input images and overfitting for small datasets. Various studies have attempted to address these challenges, such as introducing new architectures and algorithms to reduce the number of parameters and increase convergence speed, and implementing batch normalization on deep models to combat overfitting. It is rather important to consider the type and size of the dataset when choosing a CNN architecture in order to achieve optimal performance.

\subsection{Quantum Convolutional Neural Network}

The current research aims to explore the potential of Quantum Convolutional Neural Networks (QCNNs) in addressing the limitations of classical CNNs in solving quantum physics problems. The exponential growth of data size as the system size increases has been a significant hindrance in utilizing classical computing methods to solve quantum physics problems. QCNNs address this challenge by utilizing qubits to represent data in a quantum environment and applying CNN structures to quantum computers.

QCNNs are based on the fundamental concepts and structures of classical CNNs, but adapt them to the realm of quantum systems. The utilization of qubits in a quantum environment allows for the property of superposition to be utilized, where qubits can exist in multiple states at the same time. This property of superposition plays a vital role in quantum computing tasks, as it allows quantum computers to perform multiple tasks in parallel without the need for a fully parallel architecture or GPUs.

In QCNNs, the image is first encoded into a quantum circuit using a given feature map, such as Qiskit's ZFeatureMap or ZZFeatureMap. Alternating convolutional and pooling layers are then applied to the encoded image, reducing the dimensionality of the circuit until only one qubit remains. The output of this remaining qubit is measured to classify the input image. The Quantum Convolutional Layer consists of a series of two-qubit unitary operators that recognize and determine relationships between the qubits in the circuit. The Quantum Pooling Layer, however, reduces the number of qubits by performing operations on each qubit until a specific point, and then discarding certain qubits in a specific layer.

In QCNNs, each layer contains parametrized circuits, meaning that the output can be altered by adjusting the parameters of each layer. During training, these parameters are adjusted to reduce the loss function of the QCNN. The present research aims to investigate the potential of QCNNs in addressing the limitations of classical CNNs and solving quantum physics problems.

\captionsetup[figure]{labelformat=empty}
\begin{figure}[h]
\begin{center}
    \includegraphics[scale=0.7]{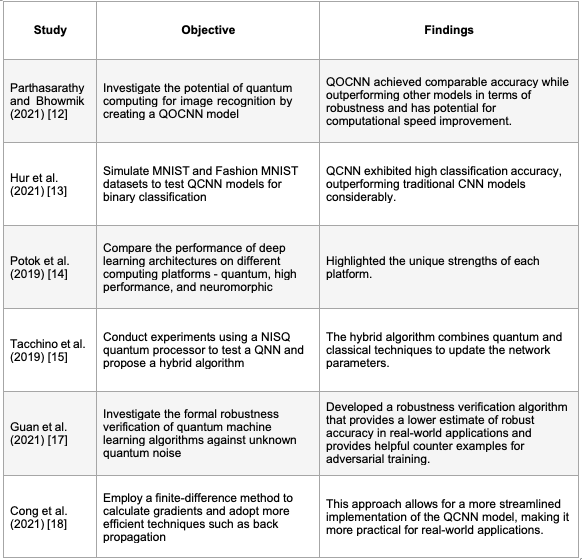}
\caption{\textbf{Table-1 } Major works in  Quantum CNN} \label{fig2}
\end{center}
\end{figure}
\captionsetup[figure]{labelformat=default}

The study by Rishab Parthasarathy and Rohan Bhowmik
\cite{lib_12} aimed to investigate the potential of quantum computing for efficient image recognition by creating and assessing a novel machine learning algorithm, the quantum optical convolutional neural network (QOCNN). The QOCNN architecture combines the quantum computing paradigm with quantum photonics and was benchmarked against competing models, achieving comparable accuracy while outperforming them in terms of robustness. Additionally, the proposed model has significant potential for computational speed improvement. The results of this study demonstrate the significant potential of quantum computing for the development of artificial intelligence and machine learning.
 
Subsequently, other studies, such as those by Tak Hur et al.  \cite{lib_13}, Potok et al.\cite{lib_14}, Tacchino et al.  \cite{lib_15}, and Ji Guan et al.  \cite{lib_16} have been conducted to explore the potential of quantum computing in machine learning.

Tak Hur et al.  \cite{lib_13} conducted a study in which they simulated the MNIST and Fashion MNIST datasets with Pennylane and various combinations of factors to test 8- qubit quantum convolutional neural network (QCNN) models for binary classification. The results of this study revealed that QCNN exhibited high classification accuracy, with the highest example being 94\% for Fashion MNIST and close to 99\% for MNIST. Furthermore, they compared the performance of QCNN to traditional convolutional neural networks (CNN) and found that, given the same training settings for both benchmarking datasets, QCNN outperformed CNN considerably.

Potok et al.  \cite{lib_14} also conducted a study that compared the performance of deep learning architectures on three different types of computing platforms - quantum, high performance, and neuromorphic, and highlighted the unique strengths of each. Tacchino et al. carried out experiments using a NISQ quantum processor to test a quantum neural network (QNN) with a small number of qubits, and proposed a hybrid algorithm that combines quantum and classical techniques to update the network parameters.

Ji Guan et al. in their study \cite{lib_17} investigated the formal robustness verification of quantum machine learning algorithms against unknown quantum noise. They discovered an analytical bound that can be efficiently calculated to provide a lower estimate of robust accuracy in real-world applications. Furthermore, they developed a robustness verification algorithm that can precisely verify the $\epsilon$-robustness of quantum machine learning algorithms and also provides helpful counter examples for adversarial training. Tensor networks are widely recognized as a powerful data structure for implementing large-scale quantum classifiers, such as QCNNs with 45 qubits in \cite{lib_18}. In order to meet the demands of NISQ devices with more than 50 qubits, the authors integrated tensor networks into their robustness verification algorithm for practical applications. However, more research is needed to fully understand the significance of robustness in quantum machine learning, particularly through more experiments on real-world applications such as learning the phases of quantum many-body systems.

Iris Cong et al. in their work \cite{lib_18} employed a finite- difference method to calculate gradients and due to the structural similarity of QCNN with its classical counterpart, they adopted more efficient techniques such as back propagation. This approach allows for a more streamlined implementation of the QCNN model, making it more practical for real-world applications.

Overall, the results of these studies demonstrate the enormous potential that quantum computing holds for the development of artificial intelligence and machine learning, specifically in terms of performance and accuracy. The QCNN \cite{lib_19,lib_20}models show promising results in terms of classification accuracy and outperforming traditional CNN models. Additionally, the comparison of deep learning architectures on different types of computing platforms highlights the unique strengths of quantum computing in this field.

\subsection{Datasets}
We are training our model on the MNIST dataset. The MNIST dataset is a widely used dataset for training and testing image recognition algorithms. It contains 60,000 training examples and 10,000 test examples of handwritten digits, each represented as a 28x28 grayscale image. The digits in the dataset have been size-normalized and centred in the image to ensure consistency.

\section{Method}
\subsection{Proposed model}
\begin{figure}[h]
\begin{center}
    \includegraphics[scale=0.9]{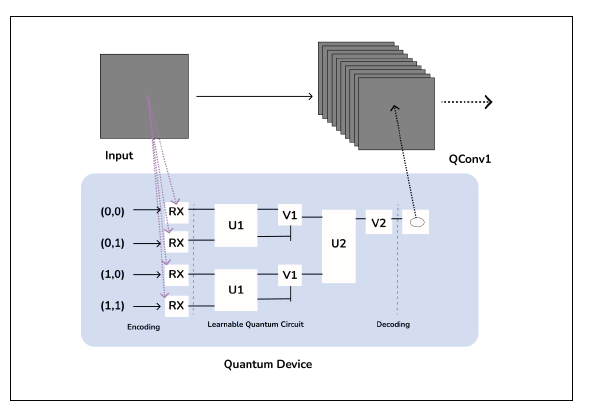}
        \caption{Proposed Quantum Neural Network model \label{fig4}}
    \end{center}
\end{figure}

Figure \ref{fig4} shows the architecture of the proposed quantum neural network model. In this proposed QCNN, the convolutional layer is modeled as a quasi-local unitary operation on the input state density. This unitary operator, denoted by Ui, is applied on several successive sets of input qubits, up to a predefined depth. The pooling layer is implemented by performing measurements on some of the qubits and applying unitary rotations Vi to the nearby qubits.

The rotation operation is determined by the observations on the qubits. After the required number of blocks of convolutional and pooling unitaries, the unitary F implements the fully connected layer. A final measurement on the output of F yields the network output. In a QCNN, the final qubit(s) is/are measured, and the measurement result is used to determine the class of the input image. The measurement result is typically a probability distribution over the possible classes. The class with the highest probability is chosen as the final output of the QCNN. The decoding process in a QCNN can be done in different ways, depending on the specific implementation of the QCNN.Here, the final qubit is measured in a computational basis, and the measurement result is used to determine the class of the input image.

Quantum Convolutional Neural Networks (QCNNs) can be mathematically modeled using quantum circuits and linear algebra. In a QCNN, the input data is represented as a quantum state, which is initialized using a set of single-qubit gates. The convolution operation in QCNNs is implemented using a set of trainable quantum filters, which are represented by unitary matrices. The pooling operation is performed using specific quantum circuits composed of quantum gates that operate on the state of the quantum register. The performance of QCNNs is evaluated using a loss function, such as the mean squared error (MSE) function. The most commonly used loss function is the mean squared error (MSE) function, which is defined as:
\begin{center}
$L(y,y_{hat})=1/n \sum (y_i - y_{hat_i})^2$
\end{center}
The optimization is performed using a quantum optimizer, such as the Variational Quantum Eigensolver (VQE), which adjusts the parameters of the quantum filters to minimize the loss function. The VQE algorithm uses the gradient descent method to minimize the loss function and update the parameters of the quantum filters. The update rule for the parameters can be represented mathematically as:

\begin{center}
$\theta_{new}=\theta_{old}-\alpha~\triangledown L(\theta_{old})$
\end{center}

The mathematical modeling of QCNNs involves the use of quantum circuits, quantum gates, quantum states, and linear algebra to perform the convolution and pooling operations and optimize the parameters of the quantum filters to minimize the loss function.\\

\textbf{Task flow}\\

\begin{figure}[h]
\begin{center}
    \includegraphics[scale=0.6]{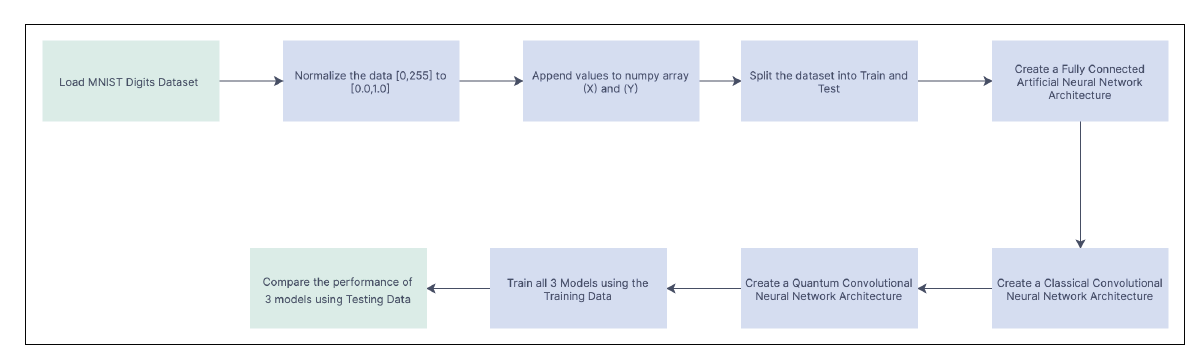}
        \caption{Task flow \label{fig5}}
    \end{center}
\end{figure}

The present research paper proposes a framework for a quantum neural network model, which is represented through a task flow diagram depicted in Figure \ref{fig5}. 
The proposed model involves the following steps: First, a standard dataset for image classification tasks is collected, in this project the MNIST dataset, consisting of images of handwritten digits, is used. Then, the algorithms to be compared are selected, in this project, an Artificial Neural Network (ANN), a Quantum Convolutional Neural Network (QCNN) and a Classical Convolutional Neural Network (CNN) are compared for their performance. Subsequently, the dataset is preprocessed to prepare it for the training of the algorithms, including scaling, normalization, and data augmentation. The models are then trained on the preprocessed dataset using an iterative process. In this process, the models are trained on batches of data, and the weights are updated based on their performance on the training data. After the models are trained, their performance is evaluated on a separate test dataset, where the accuracy and loss curves are measured to compare the performance of the ANN, QCNN and CNN models. Finally, the results are analyzed and interpreted to draw conclusions about the performance of the QCNN and CNN models for image classification tasks. Additionally, the potential advantages of QCNNs in solving complex problems using qubits are explored.

\section{Result}
The experimental results depicted in Table 2 showcase the construction of four distinct models

\begin{figure}[h]
\begin{center}
    \includegraphics[scale=1.2]{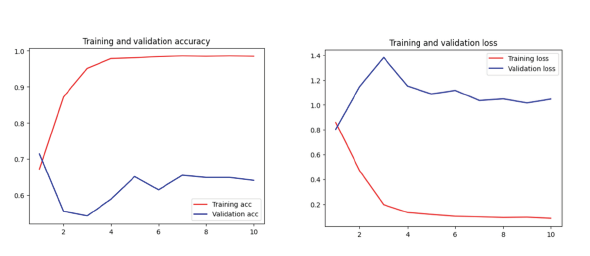}
        \caption{QNN  training vs testing (a) accuracy (b) loss \label{fig6}}
    \end{center}
\end{figure} 

\begin{figure}[h]
\begin{center}
    \includegraphics[scale=1.2]{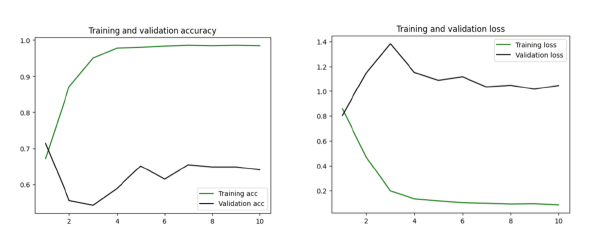}
        \caption{QCNN  training vs testing (a) accuracy (b) loss \label{fig7}}
    \end{center}
\end{figure} 

\captionsetup[figure]{labelformat=empty}
\begin{figure}[h]
\begin{center}
    \includegraphics[scale=0.7]{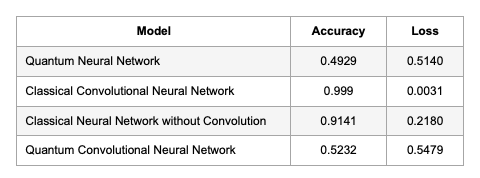}
\caption{\textbf{Table-2 }Model Comparison} \label{fig8}
\end{center}
\end{figure}
\captionsetup[figure]{labelformat=default}

used for binary classification on the MNIST dataset.  These models encompass the Quantum Neural Network (QNN), Classical Convolutional Neural Network (CNN), Classical Neural Network (NN) without Convolution, and Quantum Convolutional Neural Network (QCNN). Their training objective was to accurately classify digits as either 0 or 7, with a strong emphasis on achieving high accuracy.

Upon evaluation of a reduced-scale dataset, the classical algorithms demonstrated remarkable accuracy levels, approaching 1.0. Specifically, the CNN model achieved an accuracy of 0.999, accompanied by a low loss value of 0.0031. Similarly, the NN model achieved an accuracy of 0.9141, albeit with a slightly higher loss value of 0.2180. Contrastingly, when the quantum algorithms and models were executed using various input parameters such as batch size and epochs, the QNN model exhibited accuracy within the range of 0.50 to 0.60 and he accuracy of the QCNN model fell within the range of 0.52 to 0.61, as indicated by the accuracy curves presented in Figure \ref{fig6} and Figure \ref{fig7}.

\section{Conclusion}

\begin{figure}[h]
\begin{center}
    \includegraphics[scale=0.35]{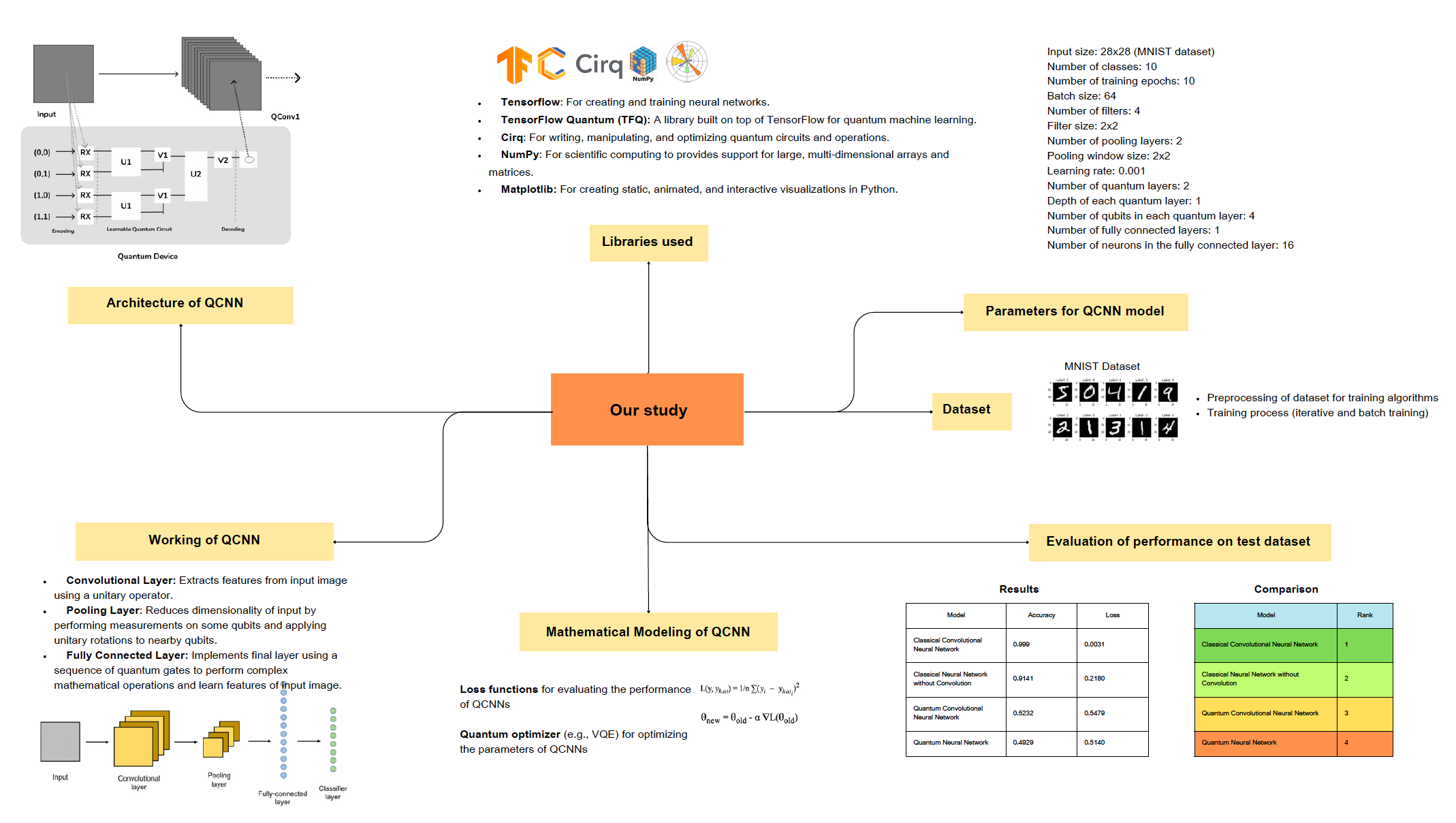}
        \caption{Mindgraph of our study \label{fig1}}
    \end{center}
\end{figure}

Figure \ref{fig1} shows the Mindgraph of the our study.  The results presented in this study demonstrate that classical neural networks outperform quantum neural networks for binary classification tasks involving the MNIST dataset. Specifically, the classical CNN and NN models achieved accuracy scores of 0.999 and 0.9141, respectively, while the accuracy scores of the quantum QNN and QCNN models were in the range of 0.5-0.6 and 0.52-0.61, respectively. \\

This finding has significant implications for the field of quantum computing and machine learning, as it suggests that classical neural networks may be more effective than quantum neural networks for certain tasks. However, it is important to note that the results presented in this study may not be generalizable to other datasets and tasks. Future research should explore the effectiveness of quantum neural networks for a wider range of tasks and datasets. In appendix \ref{appendix:a} section,  summaries a few related QCNN studies.

The small size of the MNIST dataset, containing only 60,000 training images and 10,000 test images, which, after further required preprocessing is reduced to 1000-2000 training and testing images respectively, is likely one of the reasons for the lower performance of QCNN algorithms in binary classification tasks compared to classical CNN and NN. Another potential cause for the lower performance of QCNN algorithms is their relatively new and complex architecture, which may require more optimization effort. Additionally, hardware limitations of quantum computers, such as their limited number of qubits and coherence times, could also play a role in their lower performance on small datasets like MNIST. Moreover, the sensitivity of quantum computers to noise and errors is another factor that could affect the accuracy and performance of QCNN algorithms, particularly for near-term quantum computers with high error rates. Lastly, it is possible that the MNIST dataset does not offer a clear quantum advantage over classical algorithms, and therefore, the performance of QCNN algorithms may not significantly outperform classical CNN and NN in this case.\\

\textbf{Discussions}\\

One potential avenue for future research is to investigate the effectiveness of hybrid models that combine classical and quantum neural networks. This approach has shown promise in previous studies and could potentially improve the performance of quantum neural networks for certain tasks. Additionally, the use of quantum computing hardware could potentially yield better results than simulations on classical computers.

Another area for future research is to explore the potential of quantum neural networks for unsupervised learning tasks. While classical neural networks have achieved significant success in supervised learning tasks, their effectiveness in unsupervised learning tasks is still limited. Quantum neural networks, on the other hand, have shown promise for unsupervised learning tasks such as clustering and dimensionality reduction.

In addition, the utilization of quantum computing hardware holds the potential to outperform classical computer simulations. Quantum computers excel when confronted with complex datasets and large-scale data, leading to significantly improved outcomes compared to classical computers. The inherent processing capabilities of quantum systems allow them to effectively tackle intricate computational challenges, making them highly advantageous for handling complex datasets. Consequently, quantum computing is expected to yield substantial advancements and superior results in various domains where classical computers face limitations.

In conclusion, the results of this study highlight the current limitations of quantum neural networks for binary classification tasks involving the MNIST dataset. However, further research is needed to fully explore the potential of quantum neural networks for various applications and to determine whether they can outperform classical neural networks for certain tasks.
%

\vspace{0.5cm}
\textbf{Repository} \url{https://github.com/IamRash-7/capstone_project}

\appendix
 
 \section{Appendix: Other QCNN and related models}
\label{appendix:a}
\captionsetup[figure]{labelformat=empty}
\begin{figure}[h]
\begin{center}
    \includegraphics[scale=0.67]{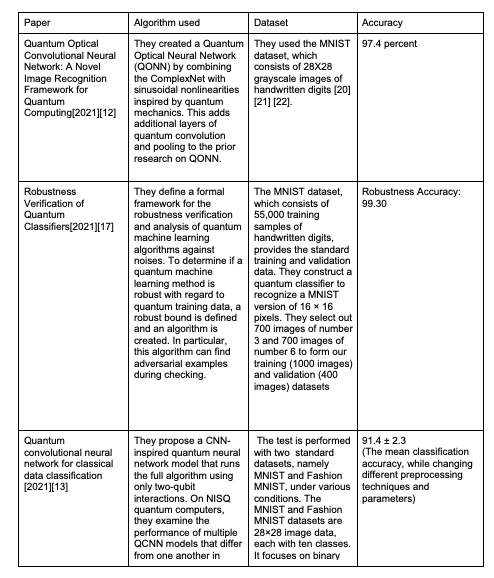}
\end{center}
\end{figure}

\newpage

\captionsetup[figure]{labelformat=empty}
\begin{figure}[h]
\begin{center}
    \includegraphics[scale=0.67]{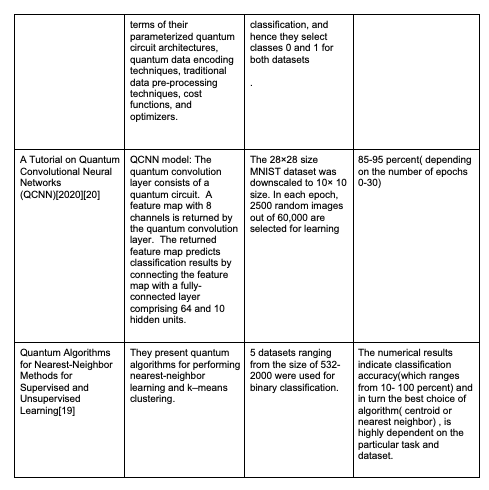}
\caption{\textbf{Table-3 }Other QCNN and related models} \label{fig9}
\end{center}
\end{figure}

\end{document}